\documentclass[sigconf,]{acmart}
\usepackage{multirow}
\usepackage{graphicx}
\usepackage{xspace}
\usepackage{amsthm}
\usepackage{enumitem}
\usepackage{booktabs}
\usepackage{float}
\usepackage{caption}
\usepackage{subcaption}
\usepackage{balance}
\usepackage{breakurl}
\usepackage[toc,page]{appendix}

\AtBeginDocument{%
  \providecommand\BibTeX{{%
    \normalfont B\kern-0.5em{\scshape i\kern-0.25em b}\kern-0.8em\TeX}}}

\setcopyright{acmcopyright}
\copyrightyear{2018}
\acmYear{2018}
\acmDOI{XXXXXXX.XXXXXXX}

\acmPrice{15.00}
\acmISBN{978-1-4503-XXXX-X/18/06}

\begin{document}
\settopmatter{printacmref=false} 
\renewcommand\footnotetextcopyrightpermission[1]{}
\title[CARec]{Collaborative Semantic Alignment in Recommendation Systems}



\author{Chen Wang}
\email{cwang266@uic.edu}

\orcid{0000-0001-5264-3305}
\affiliation{%
  \institution{University of Illinois Chicago}
  \country{USA}
}

\author{Liangwei Yang}
\email{lyang84@uic.edu}
\orcid{0000-0001-5660-766X}
\affiliation{%
  \institution{University of Illinois Chicago}
  \country{USA}
}

\author[]{Zhiwei Liu}
\email{zhiweiliu@salesforce.com}
\affiliation{%
  \institution{Salesforce AI}
  \country{USA}
}

\author[]{Xiaolong Liu}
\email{xliu262@uic.edu}

\author[]{Mingdai Yang}
\email{myang72@uic.edu}
\affiliation{
  \institution{University of Illinois Chicago}
  \country{USA}
}

\author[]{Yueqing Liang}
\email{yliang40@hawk.iit.edu}
\affiliation{
  \institution{Illinois Institute of Technology}
  \country{USA}
}

\author[]{Philip S. Yu}
\affiliation{%
  \institution{University of Illinois Chicago}
  \country{USA}}
\email{psyu@uic.edu}
\renewcommand{\shortauthors}{Chen and Liangwei, et al.}

\begin{abstract}
Traditional recommender systems primarily leverage identity-based (ID) representations for users and items, while the advent of pre-trained language models (PLMs) has introduced rich semantic modeling of item descriptions. However, PLMs often overlook the vital collaborative filtering signals, leading to challenges in merging collaborative and semantic representation spaces and fine-tuning semantic representations for better alignment with warm-start conditions. Our work introduces CARec, a cutting-edge model that integrates collaborative filtering with semantic representations, ensuring the alignment of these representations within the semantic space while retaining key semantics. Our experiments across four real-world datasets show significant performance improvements. CARec's collaborative alignment approach also extends its applicability to cold-start scenarios, where it demonstrates notable enhancements in recommendation accuracy. The code will be available upon paper acceptance.

\end{abstract}




\received{20 February 2007}
\received[revised]{12 March 2009}
\received[accepted]{5 June 2009}

\maketitle


\section{Introduction}
In the contemporary digital landscape, recommendation systems (RecSys) have emerged as indispensable tools, greatly enhancing user experiences across a wide range of online platforms, including e-commerce, content streaming, and social media networks~\cite{bonab2021crossmarket, anderson2020algorithmic}. These systems are instrumental in guiding users towards content, products, or services that align with their preferences, thus significantly contributing to user satisfaction on online platforms. Traditionally, modern recommendation models have relied on unique identifiers (IDs) to represent both users and items, transforming these IDs into embedding vectors through learnable parameters to effectively capture and predict user preferences~\cite{Collaborative_Filtering, lightgcn}.

Despite the notable success of ID-based recommendation systems (IDRec) in scenarios where sufficient user-item interaction data is available, commonly referred to as the warm setting, their dependency on historical interactions poses significant limitations. Specifically, IDRec struggle to generate reliable recommendations in situations characterized by sparse or non-existent user-item interactions, known as the cold start problem~\cite{multi-task-item-attribute-graph}. To mitigate these challenges, Semantic-Based Recommendation Models (SemRec) ~\cite{transrec,javed2021review} have been introduced that take advantage of textual content enrich the recommendation process with additional context and information about users and items. Traditionally, Semantic-Based Recommendations have been utilized mainly to tackle the cold-start problem. However, their integration into a wider array of applications has been limited due to the relative underperformance of semantic embeddings. These embeddings, often generated from word embeddings~\cite{word2vec} and basic neural network models~\cite{graph-convol-web-scale}, have not matched the effectiveness of the more sophisticated embeddings derived from item IDs.

Recent breakthroughs~\cite{bert,Sentence_bert,instructor} in pre-trained language models (PLMs) have significantly advanced the representation of textual information in the field of Natural Language Processing (NLP). These advancements present a valuable chance to enhance the efficacy of SemRec. A crucial question emerges: Can we merge IDRec with SemRec to leverage the strengths of both, thereby improving recommendations in both warm and cold settings? However, efforts to merge IDRec with SemRec through methods like addition or concatenation have not led to the anticipated improvements. ~\cite{where-to-go-next}. This may be due to the fundamental differences in their representation spaces. IDRec, which learns from user-item interactions, is considered a collaborative signal, while SemRec, derived from text through a semantic Encoder, is viewed as a semantic signal. Despite significant enhancements in semantic embeddings through modern encoders, they still fall short of the performance achieved by IDRec. This drawback could be attributed to semantic embeddings in SemRec often being pre-trained on tasks~\cite{bert,Sentence_bert} not directly related to recommendation, which, while capturing meaningful content information, fail to fully represent the specific preferences and distribution patterns of the recommendation data. This limitation is particularly evident in warm setting scenarios, where SemRec underperforms~\cite{where-to-go-next}. To successfully integrate IDRec and SemRec, we must address two main challenges: (1) bridging the gap between collaborative and semantic representation spaces and (2) refining the semantic representations to better suit warm conditions.

To address the two challenges discussed previously, we introduce a novel training strategy called \textbf{CARec} (Collaborative Alignment for Recommendation), which merges collaborative signals with semantic information in a unique way. CARec draws inspiration from Reciprocal Teaching, a teaching method where students alternate between being teachers and students, thereby enhancing their understanding through mutual exchange. Our learning framework introduces two innovative features to integrate IDRec with SemRec in a novel manner. Unlike traditional methods that merge ID and semantic information through addition or concatenation~\cite{where-to-go-next}, our framework assigns distinct roles to users and items. Specifically, users are represented by ID embeddings, which are randomly initialized due to the typical absence of direct textual information for users. Items, on the other hand, are represented by semantic embeddings, initialized through Pre-trained Language Models (PLMs) using textual data like titles, features, and descriptions.

Our training approach, termed "Reciprocal Alignment," diverges from traditional simultaneous training of user and item embeddings, which can contaminate the semantic integrity of item representations when combined with the randomly initialized user IDs. To preserve the semantic information of item embeddings, our framework employs a two-phase alignment strategy: the semantic aligning phase and the collaborative refining phase. During the semantic aligning phase, items serve as "teachers," imparting their semantic attributes to guide the recommendation process. Users, acting as "students," refine their embeddings by absorbing this semantic knowledge and reflecting on their interaction history with the items. This phase is designed to bridge the collaborative-semantic representation gap, aligning user embeddings with item semantics and safeguarding against the distortion of item semantics by random initialized user embeddings. Once user embeddings have stabilized, the roles reverse: users assume the "teacher", teaching their learned preferences and behaviors, while items, now as "students", fine-tune their representations to align with user's preferences. Crucially, item representation adjustments are mediated by an adapter to ensure the semantic features of items are preserved and enriched by collaborative feedback, thus optimizing the system for both well-established and emerging user-item interactions.

To sum up, our contributions can be outlined as follows:
\begin{enumerate}[leftmargin=*]
    \item Reciprocal alignment paradigm: We introduce a novel collaborative learning paradigm, inspired by educational peer tutoring, where users and items exchange roles, enhances recommendation quality by fostering dynamic knowledge exchange.
    \item We identify the problem, uncover the challenges and propose a feasible solution for collaborative alignment to bridge the gap between collaborative filtering and the semantic representation.
    \item Experimentally, we conduct extensive experiments on four real-world datasets under both warm and cold settings to validate the effectiveness of CARec. 
\end{enumerate}

\section{Problem Definition}

In this section, we introduce a new framework termed "Collaborative Alignment". Collaborative Alignment seeks to fuse semantic information with collaborative filtering to provide more accurate personalized recommendations. It is an inevitable problem that occurs between pre-trained language models and collaborative filtering-based recommendations.

For recommendation task, we have a set of users $\mathcal{U} = \{u_1,u_2,...,\allowbreak u_{\left | \mathcal{U} \right|}\}$, a set of items $\mathcal{I} = \{i_1,i_2,...,i_{\left | \mathcal{I} \right|}\}$ and a historical interaction matrix $\textbf{R}$ of size $\left| \mathcal{U} \right| \times \left| \mathcal{I} \right|$. By treating $\mathbf{R}$ as the adjacent matrix, we can also view the historical interaction as a user-item bipartite graph $\mathcal{G}(\mathcal{U,I,E})=\{(u,i)|u\in\mathcal{U},i\in\mathcal{I},(u,i)\in\mathcal{E}\}$, where $\mathcal{E}$ is the edge set. There is an edge $(u,i)\in\mathcal{E}$ if $\mathbf{R}_{u,i}=1$ with implicit feedback. In collaborative alignment, besides the collaborative filtering signal $\mathcal{G}(\mathcal{U,I,E})$, we also have a semantic embedding for each user/item with rich semantic information represented by $\mathbf{x}_u$ and $\mathbf{x}_i$, respectively. semantic embedding is encoded from context information with corresponding pre-trained models as illustrated in Section~\ref{sec:textual_representation}. 
Encoded from pre-trained models, semantic embedding contains rich semantic information, and collaborative alignment seeks to bridge the semantic embedding with the collaborative filtering signal to provide more accurate recommendation.

\begin{figure*}
    \centering
    \includegraphics[width=\linewidth]{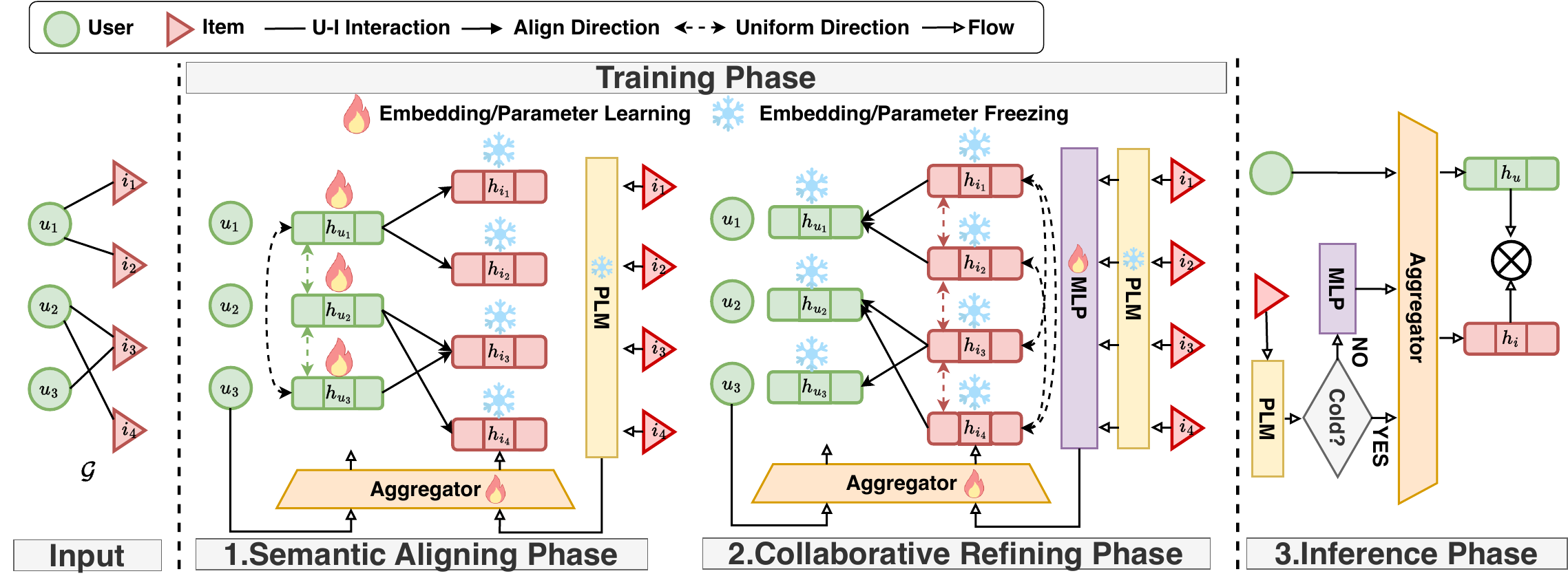}
    \caption{CARec comprises three key phases: the semantic aligning phase, the collaborative refining phase, and inference phase. During the semantic aligning phase, the model aligns user representations with the item semantic representation space. In contrast, the collaborative refining phase focuses on guiding item representations to effectively incorporate collaborative signals while preserving their semantic characteristics. Finally, in the inference Phase, the model leverages the acquired knowledge to provide personalized recommendations by utilizing the learned user embeddings and transformed item embeddings.}
    \label{fig:architec}
\end{figure*}

\section{Proposed Method}
In this section, we introduce our innovative recommendation model CARec, which is designed to address the challenge of enhancing RecSys by effectively incorporating semantic information and Collaborative Filtering (CF) signals. By leveraging historical user-item interactions, CARec learns comprehensive semantic CF-incorporated representations for both users and items. These representations not only successfully merge semantic and CF signals but also yield substantial improvements in recommendation performance, benefiting both general recommendation scenarios and challenging item cold-start scenarios. In the subsequent subsections, we detail the architecture, training phase, and inference phase of CARec, providing a comprehensive overview of our novel approach for recommendation.

\subsection{Semantic Item Representation}\label{sec:textual_representation}
To leverage the semantic-rich encoding capabilities offered by pre-trained language models (PLMs), our approach allows for using any PLM as the encoder to capture semantic item embeddings. For a given item $i$ with associated semantic features, including the item title, category, and brand, we concatenate these features into a single sentence $S_i=[w_1, w_2, ..., w_c]$, where $w$ is the text token and $c$ is the total token number. $S_i$ is then used as input to the PLM, which results in the following semantic representation for item $i$:
\begin{equation}
    \textbf{x}_i = PLM(S_i),
\end{equation}
where $\textbf{x}_i \in \mathbb{R}^{d_W}$ is $i$'s semantic embedding, and $d_W$ denotes the PLM's output embedding size. We freeze $\textbf{x}_i$ as the $0$-th layer hidden representation $\textbf{h}_i^{(0)}$ for graph convolution in Section~\ref{sec:graph_conv}.
This representation captures the rich semantic information from the item's semantic attributes, laying the foundation for the fusion of semantic and collaborative filtering signals within CARec.

\subsection{Graph Aggregator}\label{sec:graph_conv}


CARec is built upon graph aggregation to spread the rich semantic semantic embedding obtained from Section~\ref{sec:textual_representation}. With the aggregation over $\mathcal{G}(\mathcal{U,I,E})$, CARec updates user/item embedding based on the collaborative filtering signal. Let's denote the embeddings of users as $\textbf{h}_u$ and the embeddings of items as $\textbf{h}_i$, which is obtained by the graph aggregator:
\begin{equation}
    \textbf{h}_u,  \textbf{h}_i = \text{Aggregator}(\mathcal{G}(\mathcal{U,I,E}), \textbf{h}_u^{(0)}, \textbf{h}_i^{(0)}),
\end{equation}
where $\textbf{h}_u^{(0)}$ and $\textbf{h}_i^{(0)}$ represent the user/item initial embedding. $\textbf{h}_i^{(0)}$ is encoded from Section~\ref{sec:textual_representation} and $\textbf{h}_u^{(0)}$ is randomly initialized embedding due to the lack of sufficient context to encode semantic embedding. The Aggregator performs aggregation on $\mathcal{G}(\mathcal{U,I,E})$ for $K$ layers to smooth the embedding. For each layer's aggregation, the computation is defined as:
\begin{align}
    \textbf{h}_u^{(k+1)} &= \textbf{h}_u^{(k)} + \text{AGG}\left(\textbf{h}_i^{(k)} , \forall i \in \mathcal{N}(u)\right),\\
    \textbf{h}_i^{(k+1)} &= \textbf{h}_i^{(k)} + \text{AGG}\left(\textbf{h}_u^{(k)} , \forall u \in \mathcal{N}(i)\right),
\end{align}
where $\textbf{h}_u^{(k)}$ represents the embedding of user $u$ at layer $k$, and $\textbf{h}_i^{(k)}$ represents the embedding of item $i$ at layer $k$. The function $\text{AGG}$ denotes the aggregation function, which combines the embeddings of neighboring nodes. To maintain generality, we use the most widely used LGCN~\cite{lightgcn} as the aggregation function:
\begin{equation}
    \text{AGG}(\textbf{h}_i, \forall i \in \mathcal{N}(u)) = \sum_{i \in \mathcal{N}(u)} \frac{1}{\sqrt{|\mathcal{N}(u)|}\sqrt{|\mathcal{N}(i)|}} \textbf{h}_i,
\end{equation}
where $\mathcal{N}(u)$ and $\mathcal{N}(i)$ represent the set of neighboring nodes of $u$ and $i$. $\textbf{h}_i$ is the embedding of a item node $i$. User aggregation is computed in the same way.
It's worth noting that the aggregation function can be replaced with any graph aggregation function. In the subsequent section, we delve into the training phase of CARec, where we elucidate the process of learning comprehensive semantic CF-incorporated representations for users and items, a pivotal step in our innovative recommendation model.

\subsection{Training Phase}
In traditional bipartite graph learning for RecSys, a common practice involves initializing user and item ID-based embeddings randomly and updating their representations symmetrically. This symmetric learning paradigm involves aggregating the representations of neighbors for each user and item. 

However, symmetric learning, as traditionally employed, is suboptimal with semantic information. There are two primary reasons for this suboptimality. First, the initialization methods for users and items differ significantly. Users' representations are initialized as ID-based random representations, while items' representations are derived from Text-based representations generated by pre-trained language models (PLMs). This discrepancy results in two key differences: (1) \textit{Representation Space Mismatch}: The ID-based random initialization and Text-based PLM-generated representations are situated in different representation spaces, making direct symmetrical aggregation problematic. (2) \textit{Information Discrepancy}: The ID-based random initialization contains relatively meaningless information, while the Text-based PLM-generated representations incorporate influential semantic information. Combining these disparate representations in a symmetric manner can lead to issues. One critical concern is the phenomenon we term "item representation contamination." In traditional symmetric learning, during the aggregation process, user ID-based random initialized representations are aggregated to form the item representation, which is then integrated into the unified representation space. 

To address these challenges, we introduce CARec, a collaborative alignment model. CARec establishes a dynamic interchange of roles between users and items, akin to the essence of the semantic alignment phase and collaborative refining phase. Initially, items assume the role of tutors, guiding the learning process, while users adopt the role of learners, actively engaging and adapting their representations by incorporating interactions and item information to gain a deeper understanding of the system. Subsequently, the roles reverse, with users becoming tutors and items taking on the role of learners, facilitating a comprehensive knowledge exchange. 

In the following subsections, we will provide detailed insights into the semantic aligning phase and collaborative refining phase.

\subsubsection{Semantic Aligning Phase}\label{sec:item_tutoring}
In this phase, items take on the role of tutors, guiding the learning experience. On the other hand, users assume the role of learners, actively engaging and adapting their representations based on the guidance provided by the items. 

Leveraging the rich semantic information contained in item semantic representations, our objective is to align user embeddings ($\textbf{h}_u$) to the item representations ($\textbf{h}_i$) which has been shown in the Fig.~\ref{fig:architec}. To achieve this alignment and strengthen the normalized element-wise similarity between a user's representation and those of their interacted items, we employ an alignment loss inspired by the DirectAU~\cite{directau} method. The alignment loss between user $u$ and item $i$ is defined as:
\begin{equation}
    l^{\mathcal{U}}_{align} = \frac{1}{|\mathcal{E}|}\sum_{(u,i)\in \mathcal{E}}||\textbf{h}_u - \text{freeze}(\textbf{h}_i)||^2,
\end{equation}
where $\text{freeze}(\textbf{h}_i)$ indicates the frozen item embedding.

To prevent over-concentration in the representation space, the uniformity loss is also added as the regularization. In our context, we compute and apply the user uniformity loss $l^{\mathcal{U}}_{uniform}$ to optimize the learning of user representations efficiently:
\begin{equation}
    l^{\mathcal{U}}_{uniform} = log\frac{1}{|\mathcal{U}|^2}\sum_{u\in \mathcal{U}}\sum_{u*\in \mathcal{U}}e^{-2||\textbf{h}_u-\textbf{h}_{u*}||}.
\end{equation}

These two loss metrics work in synergy to maintain proximities between positive instances while dispersing random instances across the hypersphere. The final loss function in the user representation learning stage is a combination of the alignment loss and the user uniformity loss:
\begin{equation}
    \mathcal{L_{\mathcal{U}}} = l^{\mathcal{U}}_{align} + l^{\mathcal{U}}_{uniform}.
\end{equation}

The phase of user representation learning concludes upon meeting the convergence criteria, which are predicated on the prediction scores achieved on the validation set. Specifically, we employ NDCG@10 as the benchmark metric, with an early stopping parameter set at 30. At this point, we presume that users have assimilated adequate knowledge from both semantic and collaborative filtering signals, given the current state of item semantic representations. Following this, the subsequent collaborative refining phase is dedicated to refining these item semantic representations.

\subsubsection{Collaborative Refining Phase}
In contrast to aligning user representations with item semantic representations to integrate collaborative and semantic signals, the collaborative refining phase introduces subtle adjustments when refining item representations. In this phase, the primary goal is to preserve item embeddings within the semantic representation space, thus retaining semantic information while incorporating collaborative signals. To achieve this, user representations remain fixed during this phase, leveraging the knowledge gained from well-learned users. Rather than directly modifying item representations, we employ an adaptor, such as Multi-Layer Perceptron (MLP) to transform them, as illustrated in Fig.~\ref{fig:architec}(b). There are two main reasons for this approach. First, item semantic representations alone often fail to capture collaborative filtering signal, and they tend to become densely clustered~\cite{VQ_Rec, zero_shot_recommender_system, UnisRec}, which can hinder recommendation performance. Second, this approach allows us to preserve informative item semantic knowledge for use in the cold setting directly. Using an MLP to adapt new item representations in the cold setting can be problematic. The primary issue is that the process involves mapping item representations before aggregation. In the warm setting, the MLP learns to adjust the representations for effective aggregation. However, in the cold setting, where new items lack prior interactions, the MLP, being a global learner trained on warm data, struggles to appropriately adjust the item semantic representations in the absence of aggregation data.


To refine item representations, we initially apply MLP, resulting in $\widetilde{\textbf{h}}^{(0)}_i = \text{MLP}(\textbf{h}^{(0)}_i)$, where $\widetilde{\textbf{h}}^{(0)}_i$ denotes the transformed item representations. Subsequently, we compute the convoluted user and item representations as follows:
\begin{equation}
\textbf{h}_u, \widetilde{\textbf{h}}_i = \text{Aggregator}(\mathcal{G}(\mathcal{U,I,E}),\textbf{h}_u^{(0)}, \widetilde{\textbf{h}}_i^{(0)}).
\end{equation}

The alignment loss employed in item representation learning mirrors the one used in user representation learning but utilizes the transformed item representation $\widetilde{\textbf{h}}_i$. The alignment loss in the collaborative refining phase is formulated as follows:
\begin{equation}
l^{\mathcal{I}}_{align} = \frac{1}{|\mathcal{E}|}\sum_{(u,i)\in \mathcal{E}}||\text{freeze}(\textbf{h}_u) - \widetilde{\textbf{h}}_i||^2,
\end{equation}
where $\text{freeze}(\textbf{h}_u)$ is the frozen user embedding to force the training on the item.
In addition to the alignment loss, we introduce a uniformity loss for items to prevent over-concentration and ensure a well-distributed representation space. This uniformity loss encourages item representations to maintain suitable distances from each other, thereby promoting diversity in the recommendation process. The uniformity loss for items is defined as follows:
\begin{equation}
l^{\mathcal{I}}_{uniform} = \log\frac{1}{|\mathcal{I}|^2}\sum_{i\in \mathcal{I}}\sum_{i*\in \mathcal{I}}e^{-2||\widetilde{\textbf{h}}_i-\widetilde{\textbf{h}}_{i*}||}.
\end{equation}

It fosters the even distribution of item representations within the hypersphere, thereby enhancing the model's ability to capture nuanced differences between items with similar semantic features.

The final loss function for item representation learning is a combination of the alignment loss and the item uniformity loss:
\begin{equation}
\mathcal{L}_{\mathcal{I}} = l^{\mathcal{I}}_{align} + l^{\mathcal{I}}_{uniform}.
\end{equation}

Through this approach, we ensure that item representations capture both semantic information and collaborative filtering signals, leading to improved recommendation quality while retaining the flexibility to address cold-start problems.

\subsection{Inference Phase}
During the inference phase, CollabFusion leverages the learned user embeddings and transformed item embeddings to make personalized recommendations. The recommendation score $s(u, i)$ for a user-item pair $(u, i)$ is determined through the dot product between the user embedding $\textbf{h}_u$ and the transformed item embedding $\widetilde{\textbf{h}}_i$:
\begin{equation}
    s(u, i) = \textbf{h}_u^T \cdot \widetilde{\textbf{h}}_i.
\end{equation}

In the cold setting, we using the dot product between the user embedding $\textbf{h}_u$ and the item semantic embedding $\textbf{h}_i$ to calculate the recommendation score.







\section{Experiments}




This section empirically evaluates the proposed CARec on three real-world datasets. The goal is to answer the four following research questions (RQs). 
\begin{itemize}[leftmargin=*]
    \item \textbf{RQ1:} What is the performance of CARec ? 
    \item \textbf{RQ2:} Does CARec still achieve the best in the challenging cold-start recommendation ?
    \item \textbf{RQ3:} How do different parts affect CARec ?
    \item \textbf{RQ4:} Can CARec really keep the rich semantic semanticized information ? 
\end{itemize}
The detailed experiment setting and more experiment results are further illustrated in the Appendix.

\subsection{Dataset}

\begin{table}[]
\centering
\caption{The Statistics of Preprocessed Datasets: $\textbf{'Avg.U'}$ represents the average number of interactions per user, $\textbf{'Avg.I'}$ signifies the average number of interactions per item, and 'Cold-Items' indicates the count of newly introduced items.}
\label{tab:dataset}
\resizebox{\linewidth}{!}
{%
\begin{tabular}{lllll}
\toprule
 & Electronic & Office Products & Gourmet Food & Yelp\\ \hline
\textbf{\#Users} & 81,512 & 51,493 & 66,268 & 65,870\\
\textbf{\#Items} & 32,424 & 16,920 & 24,636 & 43,215\\
\textbf{\#Inters} & 623,896 & 212,795 & 307,617 & 831,470\\
\textbf{\#Avg.U} & 7.653 & 4.133 & 4.642 & 12.623\\
\textbf{\#Avg.I} & 18.232 & 11.950 & 11.857 & 19.240 \\ \hline
\textbf{\#Cold-Items} & 1,797 & 888 & 1,310 & 1,714\\
\bottomrule
\end{tabular}%
}
\end{table}

To rigorously evaluate the performance of our proposed methodology, we conduct experiments in both warm and cold settings. Key statistics of the preprocessed datasets are summarized in Table~\ref{tab:dataset}. Specifically, we use four publicly available real-world datasets from the Amazon Review Dataset\footnote{https://jmcauley.ucsd.edu/data/amazon/}: Electronics, Office Products, and Grocery and Gourmet Food and Yelp Dataset\footnote{https://www.yelp.com/dataset}. These datasets have been widely employed in prior recommendation system studies~\cite{UnisRec, VQ_Rec, cw:pre-training-graph-neural-network}. To ensure the validity of both warm and cold settings, we apply meticulous preprocessing to these datasets. Initially, in alignment with previous work~\cite{UnisRec}, we employ a 5-core filtering strategy, eliminating users and items with insufficient interactions. Subsequently, we randomly select 5\% of items to serve as the cold-start items, excising all corresponding interactions from the preprocessed dataset to create a cold-start item dataset. This ensures that cold-start items are only encountered during the testing phase. The remaining dataset is partitioned into training, validation, and testing subsets using an 80\%-10\%-10\% split. For the semantic features of items, we aggregate information from fields such as \textit{title}, \textit{categories}, and \textit{brand} in the Amazon dataset, truncating any item text exceeding 512 tokens. To prepare the Yelp dataset for analysis, we started by removing any items that did not have textual information. We then focused on interactions with ratings of 3 or higher. From the filtered set, we designated 5\% of the items as 'cold items' to create a specialized dataset for evaluating cold-start performance. We further narrowed down the dataset to include only those items and users involved in at least 15 interactions each, ensuring a more focused and relevant dataset for analysis. The remaining data was then split randomly: 80\% was used for training the model, and the remaining 20\% was equally divided between validation and testing purposes. The Yelp dataset, with its combination of business IDs (representing items) and rich textual descriptions (such as categories), provides an excellent opportunity to evaluate our model across a wide range of scenarios.

\subsection{RQ1: Evaluation of the recommendations} \label{sec:contentional_recommendation_performance}

\begin{table*}[]
\centering
\caption{Warm Setting Comparison Table. The best and second-best results are bold and underlined, respectively.  ${_\text{ID}}$ indicates ID-based models, and ${_\text{TEXT}}$ denotes models that employ item semantic representation for item embedding initialization.}
\label{tab:experiments-results}
\resizebox{\linewidth}{!}
{%
\begin{tabular}{lllllllllllllllll}
\toprule
\multicolumn{1}{c}{} & \multicolumn{4}{c}{Electronic} & \multicolumn{4}{c}{Office Products} & \multicolumn{4}{c}{Grocery and Gourmet Food} & \multicolumn{4}{c}{Yelp}\\ \hline
\multicolumn{1}{l|}{Model} & R@10 & R@50 & N@10 & \multicolumn{1}{l|}{N@50} & R@10 & R@50 & N@10 & \multicolumn{1}{l|}{N@50} & R@10 & R@50 & N@10 & \multicolumn{1}{l|}{N@50} & R@10 & R@50 & N@10 & N@50  \\ \hline
\multicolumn{1}{l|}{$\text{NeuMF}_{\text{ID}}$} & 0.0513 & 0.0675 & 0.0298 & \multicolumn{1}{l|}{0.0339} & 0.1599 & 0.1868 & 0.1051 & \multicolumn{1}{l|}{0.1113} & 0.1390 & 0.1644 & 0.0892 & \multicolumn{1}{l|}{0.0951} & 0.0469 & 0.1326 & 0.0277 & 0.0498\\
\multicolumn{1}{l|}{$\text{DirectAU}_{\text{ID}}$} & 0.0546 & 0.0694 & 0.0292 & \multicolumn{1}{l|}{0.0329} & 0.1661 & 0.1965 & 0.1000 & \multicolumn{1}{l|}{0.1069} & 0.1438 & 0.1707 & 0.0844 & \multicolumn{1}{l|}{0.0907} & 0.0467 & 0.1440 & 0.0277 & 0.0529 \\
\multicolumn{1}{l|}{$\text{LightGCN}_{\text{ID}}$} & 0.0530 & 0.0857 & 0.0331 & \multicolumn{1}{l|}{0.0413} & 0.1782 & 0.2177 & 0.1217 & \multicolumn{1}{l|}{0.1307} & 0.1526 & 0.1917 & 0.1024 & \multicolumn{1}{l|}{0.1114} & 0.0498 & 0.1441 & 0.0291 & 0.0535 \\
\multicolumn{1}{l|}{$\text{SimpleX}_{\text{ID}}$} & 0.0558 & \underline{0.1060} & 0.0305 & \multicolumn{1}{l|}{0.0429} & 0.1727 & 0.2172 & 0.1091 & \multicolumn{1}{l|}{0.1192} & 0.1519 & 0.1989 & 0.0920 & \multicolumn{1}{l|}{0.1028} & 0.0458 & 0.1355 & 0.0277 & 0.0508\\
\multicolumn{1}{l|}{$\text{NCL}_{\text{ID}}$} & \underline{0.0564} & 0.1008 & \underline{0.0354} & \multicolumn{1}{l|}{\underline{0.0464}} & \underline{0.1841} & \underline{0.2304} & \underline{0.1262} & \multicolumn{1}{l|}{\underline{0.1368}} & \underline{0.1585} & 0.2063 & \underline{0.1074} & \multicolumn{1}{l|}{\underline{0.1183}} & \underline{0.0524} & \underline{0.1570} & \underline{0.0317} & \underline{0.0583}\\ \hline
\multicolumn{1}{l|}{\textit{Instructor-xl}} & 0.0108 & 0.0255 & 0.0061 & \multicolumn{1}{l|}{0.0094} & 0.0028 & 0.0121 & 0.0012 & \multicolumn{1}{l|}{0.0031} & 0.0016 & 0.0045 & 0.0012 & \multicolumn{1}{l|}{0.0018} & 0.0008 & 0.0029 & 0.0004 & 0.0010\\
\multicolumn{1}{l|}{$\text{NeuMF}_{\text{TEXT}}$} & 0.0389 & 0.0540 & 0.0237 & \multicolumn{1}{l|}{0.0275} & 0.1271 & 0.1579 & 0.0846 & \multicolumn{1}{l|}{0.0916} & 0.1166 & 0.1417 & 0.0736 & \multicolumn{1}{l|}{0.0795} & 0.0255 & 0.0835 & 0.0151 & 0.0301\\
\multicolumn{1}{l|}{$\text{DirectAU}_{\text{TEXT}}$} & 0.0551 & 0.0718 & 0.0294 & \multicolumn{1}{l|}{0.0336} & 0.1671 & 0.1963 & 0.1004 & \multicolumn{1}{l|}{0.1071} & 0.1434 & 0.1703 & 0.0843 & \multicolumn{1}{l|}{0.0906} & 0.0472 & 0.1430 & 0.0278 & 0.0526\\
\multicolumn{1}{l|}{$\text{LightGCN}_{\text{TEXT}}$} & 0.0560 & 0.0909 & 0.0320 & \multicolumn{1}{l|}{0.0407} & 0.1782 & 0.2177 & 0.1217 & \multicolumn{1}{l|}{0.1307} & 0.1525 & 0.1965 & 0.1030 & \multicolumn{1}{l|}{0.1131} & 0.0496 & 0.1433 & 0.0290 & 0.0533\\
\multicolumn{1}{l|}{$\text{SimpleX}_{\text{TEXT}}$} & 0.0514 & 0.0931 & 0.0334 & \multicolumn{1}{l|}{0.0437} & 0.1759 & 0.2121 & 0.1202 & \multicolumn{1}{l|}{0.1284} & 0.1513 & 0.1891 & 0.1015 & \multicolumn{1}{l|}{0.1102} & 0.0495 & 0.1405 & 0.0294 & 0.0531\\
\multicolumn{1}{l|}{$\text{NCL}_{\text{TEXT}}$} & 0.0553 & 0.0929 & 0.0324 & \multicolumn{1}{l|}{0.0417} & 0.1767 & 0.2213 & 0.1174 & \multicolumn{1}{l|}{0.1275} & 0.1547 & \underline{0.2074} & 0.1021 & \multicolumn{1}{l|}{0.1142} & 0.0068 & 0.0264 & 0.0042 & 0.0093 \\ \hline
\multicolumn{1}{l|}{Wide\&Deep} & 0.0138 & 0.0485 & 0.0074 & \multicolumn{1}{l|}{0.0159} & 0.1014 & 0.1428 & 0.0609 & \multicolumn{1}{l|}{0.0705} & 0.0923 & 0.1303 & 0.0554 & \multicolumn{1}{l|}{0.0642} & 0.0219 & 0.0756 & 0.0128 & 0.0266\\
\multicolumn{1}{l|}{DCNV2} & 0.0373 & 0.0575 & 0.0216 & \multicolumn{1}{l|}{0.0266} & 0.1292 & 0.1648 & 0.0829 & \multicolumn{1}{l|}{0.0910} & 0.1160 & 0.1516 & 0.0717 & \multicolumn{1}{l|}{0.0799} & 0.0248 & 0.0837 & 0.0145 & 0.0296\\ \hline
\multicolumn{1}{l|}{CARec} & \textbf{0.0639} & \textbf{0.1069} & \textbf{0.0417} & \multicolumn{1}{l|}{\textbf{0.0523}} & \textbf{0.1878} &\textbf{ 0.2315} & \textbf{0.1312} & \multicolumn{1}{l|}{\textbf{0.1393}} & \textbf{0.1626} & \textbf{0.2098} & \textbf{0.1173} & \multicolumn{1}{l|}{\textbf{0.1267}} &\textbf{0.0578} & \textbf{0.1720} & \textbf{0.0332} & \textbf{0.0627}\\ 

\bottomrule
\end{tabular}%
}
\end{table*}

We evaluate the performance of our proposed method against various baseline approaches across four distinct datasets, with results detailed in Table~\ref{tab:experiments-results}. For the sake of clarity and comparative analysis, we categorize the baseline methods into three distinct classes: ID-based (denoted as $_{\text{ID}}$), Text-based (denoted as $_{\text{TEXT}}$), and content-based, which includes models such as Wide\&Deep and DCNV2. In ID-based models, the embedding for both the user and item is learned exclusively from their respective IDs and interactions. Text-based models, on the other hand, utilize semantic representations of items to initialize item embeddings. These semantic representations are obtained from the \textit{instructor-xl}~\cite{instructor} model. Additionally, for the Text-based baseline, we compute user representations by averaging the embeddings of items they have interacted with, as generated by \textit{instructor-xl}, and then use a dot product with item semantic representations for making recommendations. For content-based models, on the other hand, incorporate item semantic representation as features for enhanced semantic understanding.

In a comparative analysis with established baseline methods, our proposed CARec model consistently demonstrates superior performance over both ID-based and Text enhanced algorithms across a wide range of evaluation metrics. It validates that CARec can effectively incorporate collaborative filtering signal and semantic information to set the state-of-the-art performance.
By utilizing item semantic representation, different baseline methods behave differently. DirectAU, LightGCN, and SimpleX only exhibit very marginal gains compared with their ID-based counterparts. In contrast, NeuMF and NCL experience significant performance degradation when initialized with item semantic representations. It shows previous methods are unable to incorporate both collaborative filtering signal and the encoded semantic embedding from PLM effectively.

\subsection{RQ2: Cold-start Evaluation} \label{sec:cold-start_recommendation_performance}
\begin{table*}[]
\centering
\caption{Cold Setting Comparison Table. Notations consistent with the warm setting comparison.}
\label{tab:cold-start-exp}
\resizebox{\textwidth}{!}{%
\begin{tabular}{l|llllllllllllllll}
\toprule
\multicolumn{1}{c}{} & \multicolumn{4}{c}{Electronic} & \multicolumn{4}{c}{Office Products} & \multicolumn{4}{c}{Grocery and Gourmet Food} & \multicolumn{4}{c}{Yelp}\\ \hline
\multicolumn{1}{l|}{Model} & R@10 & R@50 & N@10 & \multicolumn{1}{l|}{N@50} & R@10 & R@50 & N@10 & \multicolumn{1}{l|}{N@50} & R@10 & R@50 & N@10 & \multicolumn{1}{l|}{N@50} & R@10 & R@50 & N@10 & N@50  \\ \hline
\multicolumn{1}{l|}{\textit{Instructor-xl}} & 0.0488 & 0.1197 & 0.0297 & \multicolumn{1}{l|}{0.0456} & 0.0374 & 0.1401 & 0.0159 & \multicolumn{1}{l|}{0.0382} & 0.0091 & 0.0595 & 0.0053 & \multicolumn{1}{l|}{0.0158} & 0.0067 & 0.0343 & 0.0037 & 0.0108\\
$\text{NeuMF}_{\text{TEXT}}$ & 0.0046 & 0.0250 & 0.0021 & \multicolumn{1}{l|}{0.0066} & 0.0101 & 0.0558 & 0.0047 & \multicolumn{1}{l|}{0.0145} & 0.0074 & 0.0413 & 0.0034 & \multicolumn{1}{l|}{0.0107} & 0.0030 & 0.0220 & 0.0014 & 0.0058 \\
$\text{DirectAU}_{\text{TEXT}}$ & 0.0048 & 0.0277 & 0.0023 & \multicolumn{1}{l|}{0.0073} & 0.0118 & 0.0511 & 0.0057 & \multicolumn{1}{l|}{0.0143} & 0.0082 & 0.0429 & 0.0038 & \multicolumn{1}{l|}{0.0113} & 0.0049 & 0.0229 & 0.0026 & 0.0068\\
$\text{LightGCN}_{\text{TEXT}}$ & 0.0104 & 0.0443 & 0.0049 & \multicolumn{1}{l|}{0.0124} & 0.0309 & 0.0960 & 0.0173 & \multicolumn{1}{l|}{0.0315} & 0.0119 & 0.0522 & 0.0053 & \multicolumn{1}{l|}{0.0140} & 0.0047 & 0.0201 & 0.0023 & 0.0059\\
$\text{SimpleX}_{\text{TEXT}}$ & 0.0098 & 0.0419 & 0.0049 & \multicolumn{1}{l|}{0.0120} & 0.0122 & 0.0606 & 0.0064 & \multicolumn{1}{l|}{0.0168} & 0.0113 & 0.0583 & 0.0056 & \multicolumn{1}{l|}{0.0158} & \underline{0.0260} & \underline{0.0801} & \underline{0.0144} & \underline{0.0272}\\
$\text{NCL}_{\text{TEXT}}$ & 0.0188 & 0.0716 & 0.0092 & \multicolumn{1}{l|}{0.0209} & 0.0338 & 0.1150 & 0.0143 & \multicolumn{1}{l|}{0.0320} & 0.0222 & 0.0778 & 0.0128 & \multicolumn{1}{l|}{0.0248} & 0.0051 & 0.0275 & 0.0023 & 0.0076\\ \hline
Wide\&Deep & 0.0038 & 0.0204 & 0.0018 & \multicolumn{1}{l|}{0.0055} & 0.0140 & 0.1118 & 0.0060 & \multicolumn{1}{l|}{0.0267} & 0.0150 & 0.1224 & 0.0072 & \multicolumn{1}{l|}{0.0297} & 0.0042 & 0.0260 & 0.0021 & 0.0071\\
DCNV2 & 0.0057 & 0.0282 & 0.0029 & \multicolumn{1}{l|}{0.0078} & 0.0107 & 0.0510 & 0.0047 & \multicolumn{1}{l|}{0.0134} & 0.0082 & 0.0356 & 0.0039 & \multicolumn{1}{l|}{0.0099} & 0.0045 & 0.0265 & 0.0023 & 0.0074 \\ \hline
DroupoutNet & \underline{0.0573} & \underline{0.1203} & \underline{0.0354} & \multicolumn{1}{l|}{\underline{0.0497}} & \underline{0.1313} & \underline{0.1925} & \underline{0.0844} & \multicolumn{1}{l|}{\underline{0.0982}} & \underline{0.1470} & \underline{0.2446} & \underline{0.0989} & \multicolumn{1}{l|}{\underline{0.1208}} & 0.0239 & 0.0798 & 0.0114 & 0.0241\\
Heater & 0.0036 & 0.0293 & 0.0017 & \multicolumn{1}{l|}{0.0073} & 0.0078 & 0.0412 & 0.0034 & \multicolumn{1}{l|}{0.0105} & 0.0271 & 0.0500 & 0.0095 & \multicolumn{1}{l|}{0.0146} & 0.0052 & 0.0269 & 0.0041 & 0.0074 \\ \hline
\multicolumn{1}{l|}{CARec} & \textbf{0.0617} & \textbf{0.1498} & \textbf{0.0379} & \multicolumn{1}{l|}{\textbf{0.0578}} & \textbf{0.1471} & \textbf{0.2696} & \textbf{0.0948} & \multicolumn{1}{l|}{\textbf{0.1220}} & \textbf{0.1774} & \textbf{0.3026} & \textbf{0.1226} & \multicolumn{1}{l|}{\textbf{0.1507}} & \textbf{0.0278} & \textbf{0.0842} & \textbf{0.0155} & \textbf{0.0288}\\ 

\bottomrule
\end{tabular}%
}
\end{table*}

Table~\ref{tab:cold-start-exp} provides an overview of the results of cold-start item recommendations. Key findings include: 1) Our model, CARec, consistently outperforms the best-performing baseline models in cold-start item recommendations. In contrast, DropoutNet and Heater, despite utilizing the same item semantic representations as item features, and employing pre-trained Matrix Factorization (MF) to initialize user and item representations, fall short in comparison to our model. This underscores the efficacy of CARec in leveraging item semantic knowledge and highlights the benefits of its unified representational space, seamlessly integrated into the item semantic representation space. This integration serves a dual purpose: it retains the semantic richness of item semantic descriptions while effectively addressing the cold-start item challenge. Importantly, CARec accomplishes this without the need for auxiliary modules or additional steps. 2) collaborative-based models struggle to harness the informative item semantic representation for addressing the cold-start item problem. This challenge arises from the Semantic Disparity between semantic and Collaborative spaces. Treating users and items as equal entities during training leads to an alignment of the unified representation space into a new space that diverges from the original item semantic representation space. Consequently, when a new item is introduced, its semantic representation deviates from the unified representation space, causing the model to encounter difficulties in making recommendations for the new item. 4) The performance ranking of baseline models shows notable variations in the Yelp dataset. Specifically, SimpleX$_\text{TEXT}$ outperforms DropoutNet, while NCL$_\text{TEXT}$ experiences a marked decline. This shift can be attributed to the nature of text data in the Yelp dataset, which generally consists of shorter and simpler textual content compared to the comprehensive product descriptions found in the Amazon dataset. For example, a typical Yelp entry might list a business name like "Spice Garden" along with categories such as ["Restaurants", "Food", "Seafood", "Desserts", "Indian"]. The concise and straightforward nature of Yelp's text information, unlike Amazon's detailed descriptions, impacts the effectiveness of textual embeddings. Nevertheless, our model, CARec, demonstrates superior performance by effectively integrating both ID and semantic information, maintaining its lead despite these challenges. 5) The performance of \textit{Instructor-xl} on the Electronics and Office Products datasets highlights its effectiveness, attributing to its capacity to generate meaningful semantic representations. This underscores the significant advantages of leveraging semantic representations, particularly in scenarios characterized by sparse data or the cold-start problem.

\subsection{RQ3: In-depth Analysis} \label{sec:ablation_study}
\begin{figure}
    \begin{subfigure}{0.23\textwidth}
        \includegraphics[width=\linewidth]{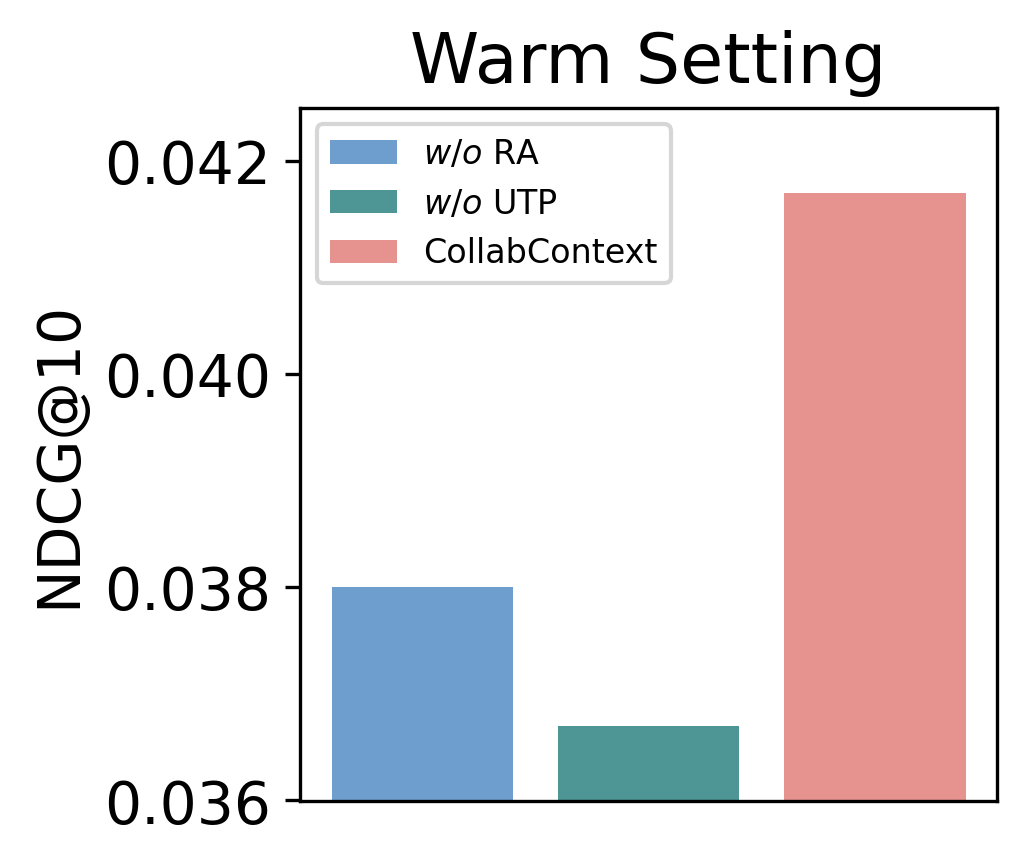}
    \end{subfigure}
    \hfill
    \begin{subfigure}{0.23\textwidth}
        \includegraphics[width=\linewidth]{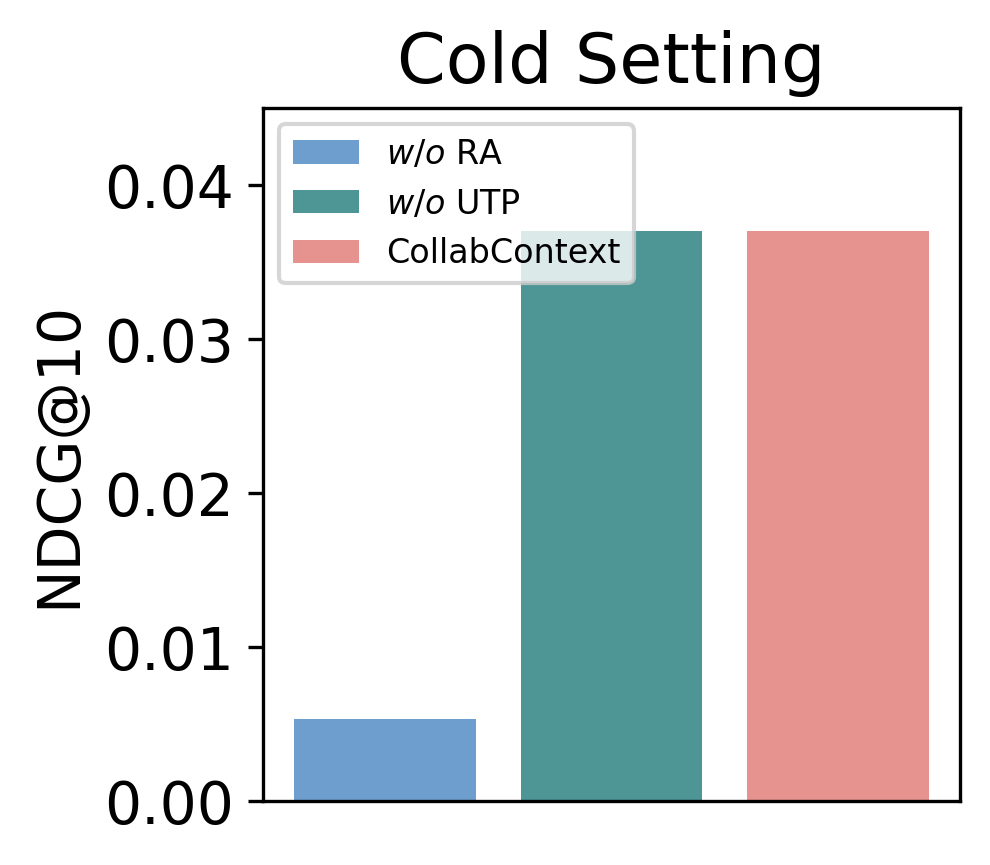}
    \end{subfigure}
    \caption{Ablation study of CARec on Electronic}
\label{fig:ablation_study}
\end{figure}

\subsubsection{Ablation Study}\label{sec:ablation-study}
In this subsection, we present a comprehensive analysis of the impact of each proposed technique and component on both the warm and cold setting performances. To facilitate a thorough comparison, we prepare two variants of the CARec model: (1) a variant without Reciprocal Alignment (RA), denoted as \underline{$w/o$ RA}, maintaining the training strategy consistent with collaborative models; and (2) a variant omits the collaborative refining phase, applying only item semantic representation.

The results of this ablation study are illustrated in Fig.~\ref{fig:ablation_study}. Notably, the absence of Reciprocal Alignment (RA), as observed in \underline{$w/o$ RA}, leads to a significant reduction in model performance. This underscores the critical importance of preserving the item semantic representation space. Additionally, the exclusion of the collaborative refining phase (UTP) reveals interesting insights. In the warm setting, omitting the UTP harms performance, as item semantic representations alone may tend to crowd together. Conversely, in the cold setting, item semantic representations still retain valuable context features that assist in addressing the cold-start challenge.

\subsubsection{Impact of Diverse Pretrained Language Models (PLMs)} \label{sec:plm}
Numerous robust Pretrained Language Models (PLMs) hold the potential to enhance RecSys by providing valuable item semantic representations. To identify the most effective PLMs for our specific objective, we conducted an evaluation of item semantic representations generated by four additional PLMs. These PLMs have achieved varying rankings on the Massive Text Embedding Benchmark (MTEB) leaderboard~\footnote{https://huggingface.co/spaces/mteb/leaderboard}, highlighting their diverse capabilities and potential contributions to the enhancement of RecSys. The five PLMs evaluated include \textit{instructor-xl}\cite{instructor}, \textit{all-MiniLM-L6-v2}\cite{Sentence_bert}, \textit{all-mpnet-base-v2}\cite{Sentence_bert}, and \textit{bge-base-en-v1.5}\cite{bge_embedding} and \textit{bert-base-unchased}\cite{bert}. The experimental results are presented in Table~\ref{tab:plms-exp}, demonstrating significant improvements in the cold setting than the default model. This underscores the vital importance of aligning the user representation space with the item semantic representation space. Notably, \textit{instructor-xl} emerges as the top-performing PLM overall, as it can generate text embeddings simply by providing the task instruction, without requiring fine-tuning. For our experiments, we used the instruction "Represent the Amazon title:" with \textit{instructor-xl} to generate the text embeddings.

\begin{table}[]
\centering
\caption{Comparison Table of PLMs. Notations consistent with warm setting comparison.}
\label{tab:plms-exp}{}{%
\begin{tabular}{l|ll|ll}
\toprule
Electronic & \multicolumn{2}{c}{Warm Setting} & \multicolumn{2}{c}{Cold Setting} \\ \hline
PLMs & R@10 & N@10 & R@10 & N@10 \\ \hline
\textit{instructor-xl} & \underline{0.0639} & \underline{0.0417} & \textbf{0.0612} & \textbf{0.0370} \\
\textit{all-MiniLM-L6-v2} & 0.0633 & 0.0414 & \underline{0.0563} & \underline{0.0343} \\
\textit{all-mpnet-base-v2} & \textbf{0.0641} & \textbf{0.0419} & 0.0556 & 0.0341 \\
\textit{bge-base-en-v1.5} & 0.0636 & 0.0414 & 0.0561 & 0.0339 \\
\textit{bert-base-unchased} & 0.0631 & 0.0411 & 0.0352 & 0.0194 \\
\bottomrule
\end{tabular}%
}
\end{table}

\subsubsection{Should We Include Additional Item Tutoring and collaborative refining phases?}
\begin{figure}
    \includegraphics[width=\linewidth]{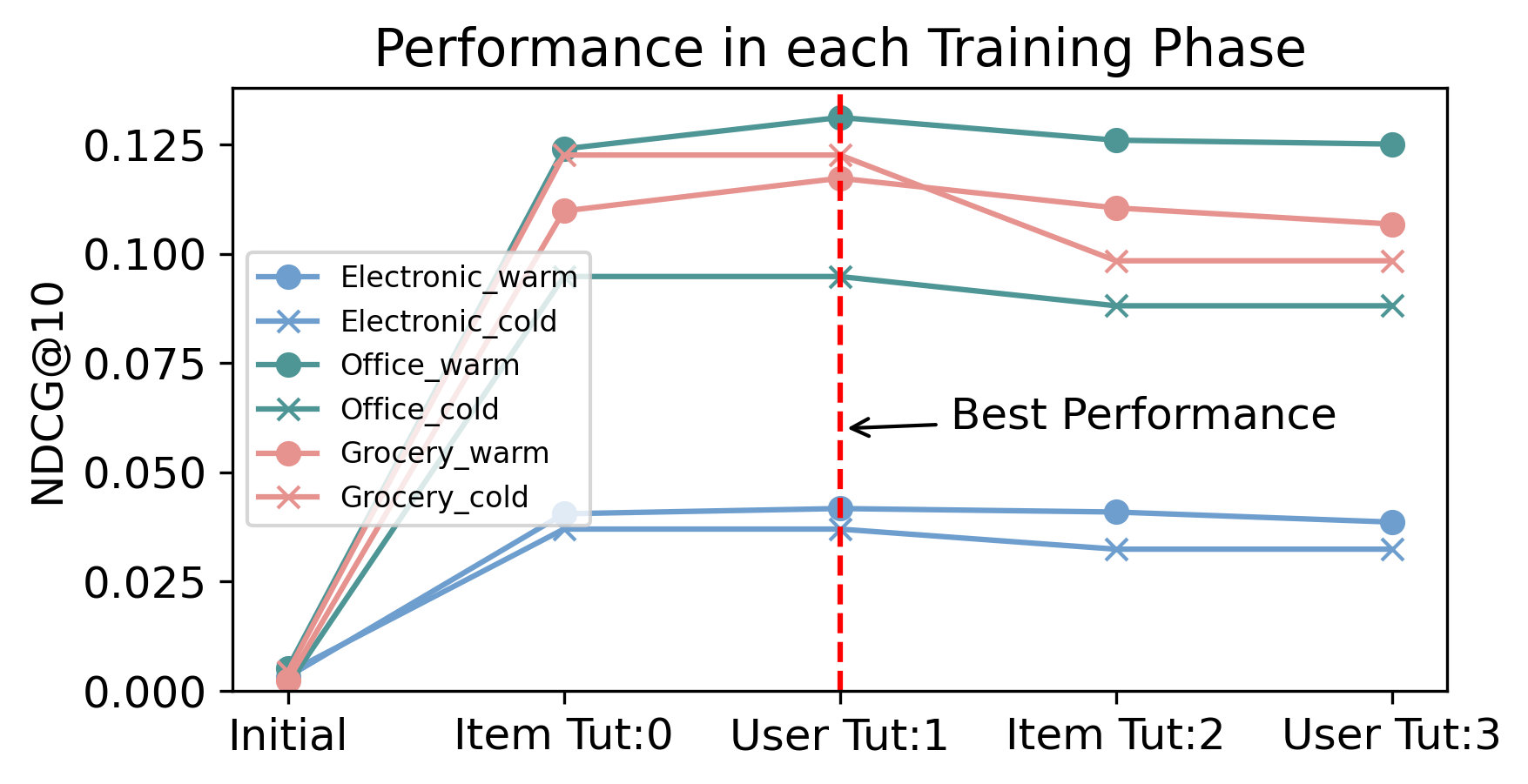}
    \caption{Overall performance in each training phase}
    \label{fig:ablation_futther_train}
\end{figure}

To investigate the advantages of further training user and item representations beyond the initial item tutoring and collaborative refining phases, we conducted experiments involving continuous learning on these representations. Fig.~\ref{fig:ablation_futther_train} presents the model's performance in both warm and cold settings across three datasets. In the plot, 'Item Tut' indicates that the current phase is the semantic aligning phase, while 'User Tut' designates the collaborative refining phase. The numbers on the x-axis represent the current training stage.

As depicted in Fig.~\ref{fig:ablation_futther_train}, CARec achieved its best performance after the first collaborative refining phase, denoted as 'User Tut:1,' across all three datasets. This suggests that user and item representations do not require additional separate training stages. Continuing to train the model beyond this point results in a performance decline, possibly due to the significant deviation of user and item representations from the item semantic representation, leading to a loss of semantic information.

\begin{table}
\caption{Impact of User Embedding Initialization on Model Performance in the Yelp Dataset.}
\label{tab:user-pooling}
\resizebox{\linewidth}{!}{%
\begin{tabular}{lllll|llll}
\hline
\multicolumn{1}{l|}{{ }} & \multicolumn{4}{c|}{Warm} & \multicolumn{4}{c}{Cold} \\ \hline
 & R@10 & R@50 & N@10 & N@50 & R@10 & R@50 & N@10 & N@50 \\ \hline
CARec & {0.0578} & { 0.1720} & { 0.0332} & {0.0627} & { \textbf{0.0278}} & { \textbf{0.0842}} & { \textbf{0.0155}} & { \textbf{0.0288}} \\ \hline
CARec$_{\text{AVG}}$ & { \textbf{0.0579}} & { \textbf{0.1722}} & { \textbf{0.0342}} & { \textbf{0.0637}} & { 0.0221} & { 0.0735} & {0.0119} & {0.0240} \\ \hline
\end{tabular}%
}
\end{table}

\subsubsection{Should users be initialized with semantic representations instead of random initialization?}  In the Yelp dataset, as illustrated in Table~\ref{tab:user-pooling}, we explore the effect of initializing user embeddings through average pooling of historical item sequences, denoted as CARec$_{\text{AVG}}$. Our findings are as follows: (1) Utilizing average pooling for initializing user embeddings enhances model performance in scenarios with abundant historical data (warm setting) but results in diminished effectiveness in data-scarce situations (cold setting). This indicates that while average pooling can somewhat narrow the representation gap, it does not offer the same level of adaptability as random initialization, especially in contexts with sparse user-item interactions. (2) Although average pooling helps bridge the initial representation gap, it does not reach the full potential of performance enhancement unless coupled with a robust training framework like ours. (3) The comparative analysis of CARec$_{\text{AVG}}$ and CARec underscores that the choice of embedding initialization strategy can significantly influence outcomes, contingent upon its integration within a systematic training methodology.

\subsubsection{Parameter Sensitivity}
\begin{figure}
    \begin{subfigure}{0.23\textwidth}
        \includegraphics[width=\linewidth]{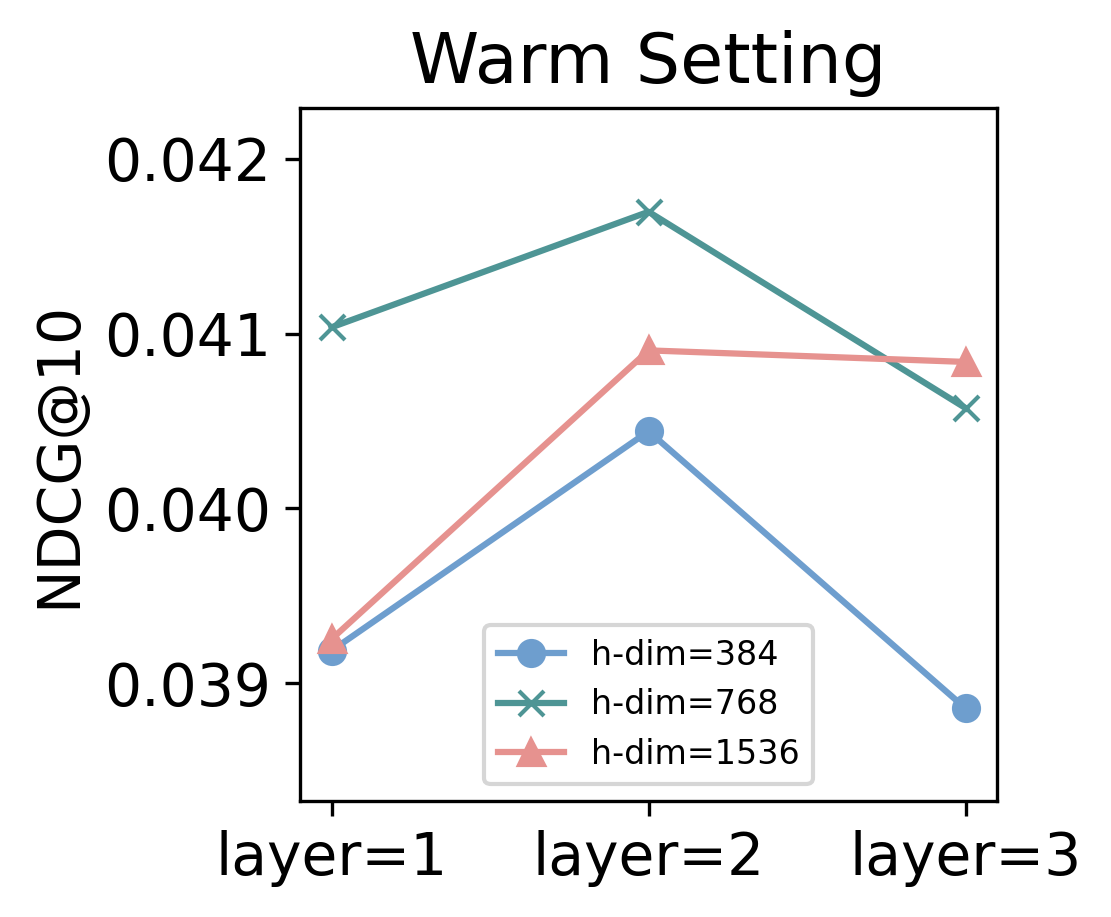}
    \end{subfigure}
    \hfill
    \begin{subfigure}{0.23\textwidth}
        \includegraphics[width=\linewidth]{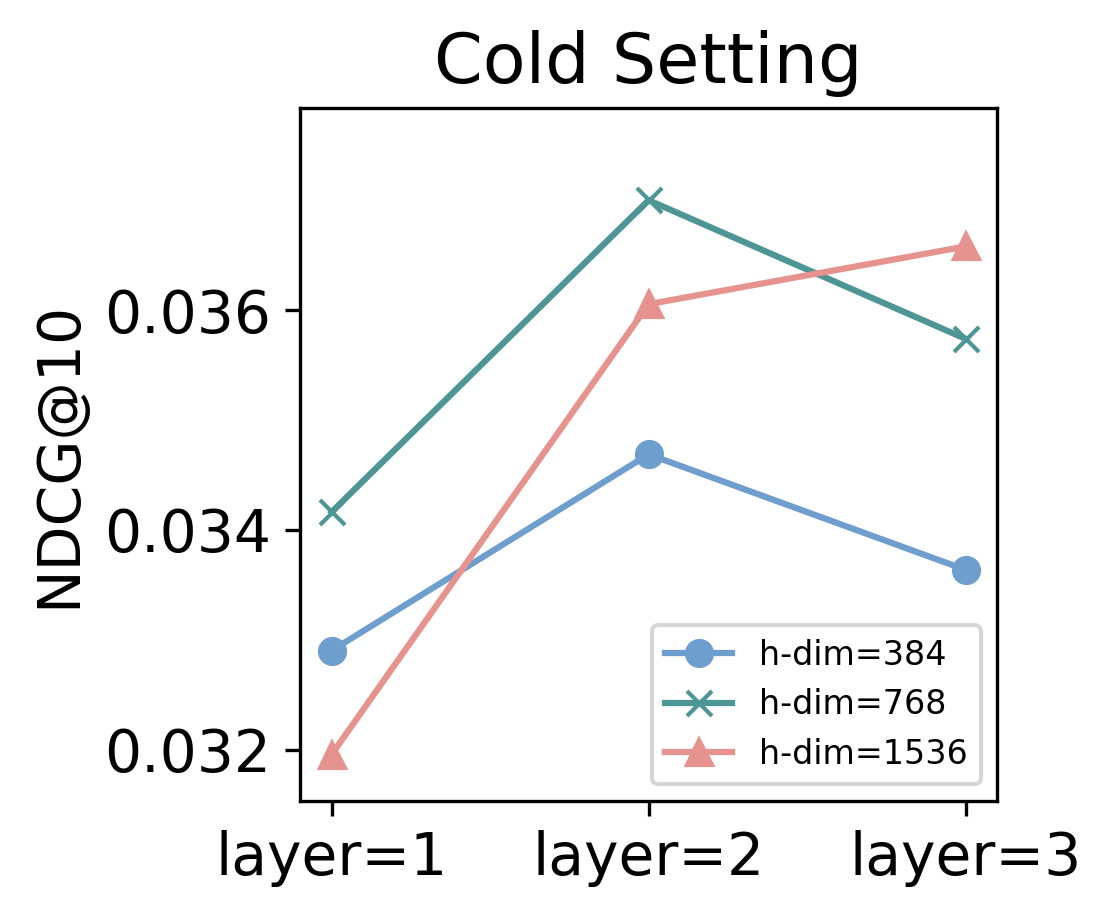}
    \end{subfigure}
    \caption{Parameter analysis of MLP on Electronic}
    \label{fig:ablation_param_study}
\end{figure}

We experimented with different configurations of the number of layers for the MLP. The sensitivity results, shown in Fig.~\ref{fig:ablation_param_study}, reveal that the layer number is not sensitive for model performance, and the highest performance is achieved when using two layers with a hidden dimension of 768.

\subsection{RQ4: Case Study}
\begin{figure}
    \begin{subfigure}{0.23\textwidth}
        \includegraphics[width=\linewidth]{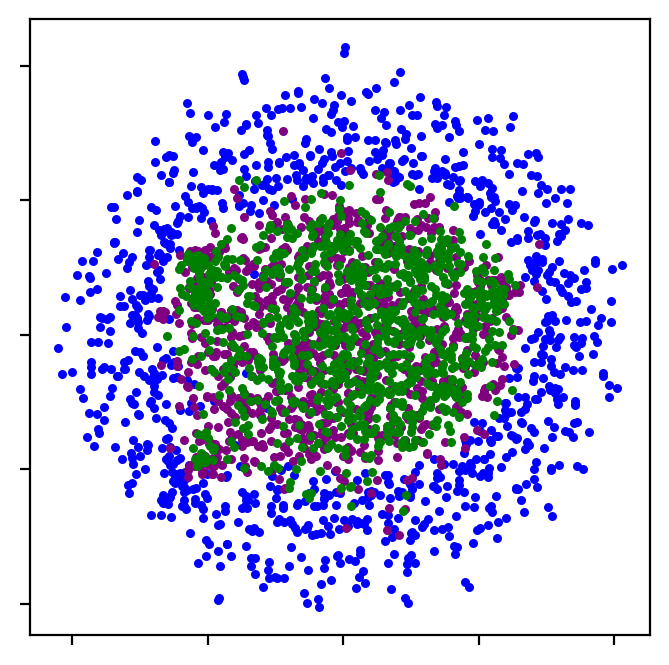}
        \caption{CF Representation Space}
    \end{subfigure}
    \hfill
    \begin{subfigure}{0.23\textwidth}
        \includegraphics[width=\linewidth]{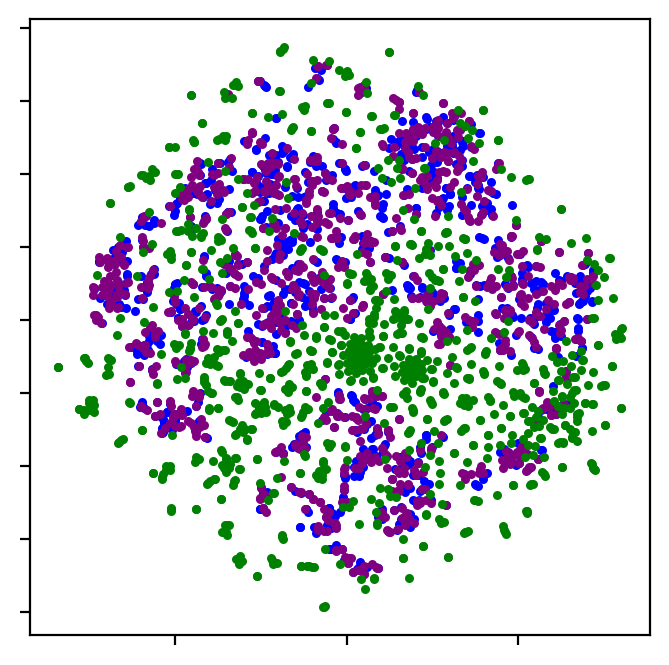}
        \caption{CC Representation Space}
    \end{subfigure}
    \caption{Comparison of representation space after model alignment. The left figure illustrates the representation space following Collaborative Filtering Alignment, while the right figure depicts the representation space after Collaborative alignment Alignment. In both figures, the \underline{blue} nodes symbolize item semantic representations, the \underline{purple} nodes represent item mapped representations by MLP, and the \underline{green} nodes denote user learned representations.}
    \label{fig:case_study}
\end{figure}

To provide visual evidence of CARec's effectiveness in aligning user representations with item semantic representations while preserving the integrity of item-learned representations, we present a case study in Fig \ref{fig:case_study}. In the left figure, which represents the representation space following traditional collaborative filtering alignment, the item semantic representation (blue) is shown surrounding the user (green) and item (purple) mapped representations by MLP. This spatial arrangement suggests that the item semantic representation is not effectively integrated into the same space as the user and item representations. In contrast, CARec, as shown in the right figure, successfully aligns the user and item learned representations within the item semantic representation space, ensuring that informative semantic information is retained. This alignment contributes to significant improvements in both warm and cold settings, showcasing the model's enhanced performance.

\section{Related work}
\subsection{Collaborative Filtering}
Collaborative Filtering (CF) is a widely used technique in modern RecSys. CF models typically represent users and items as embeddings and learn these embeddings by reconstructing historical user-item interactions. With the rise of Graph Neural Networks (GNNs)~\cite{gcn}, GNN-based RecSys have gained popularity. These methods model user-item interactions as bipartite graphs, enabling them to capture high-order connectivity. SpectralCF~\cite{spectralcf} introduced spectral convolution to improve recommendation performance, particularly for cold-start items. PinSAGE~\cite{pinsage} designed a random walk strategy for large-scale graphs, specifically on the Pinterest platform. NGCF~\cite{ngcf} aggregates information from neighboring nodes and incorporates collaborative signals into embeddings. LightGCN~\cite{lightgcn} simplified NGCF, achieving better performance and reduced training time. However, these collaborative models treat users and items equally and learn their representations simultaneously, which may not work well when incorporating informative item semantic representations.

\subsection{Cold-start Recommendation}
Addressing the cold-start problem requires bridging the gap between warm-start and cold-start items. To achieve this, side information, particularly content features, is often integrated into CF-based recommendation models. These content features serve as a link to capture the collaborative signal for cold-start items. For example, models like DropoutNET~\cite{droupoutnet} and CC-CC~\cite{cc-cc} randomly omit certain collaborative embeddings, enhancing the robustness of CF-based models while implicitly tapping into information related to the collaborative signal from item content features. In contrast, some approaches focus on explicitly modeling the correlation between content information and collaborative embeddings~\cite{heater}. In our approach, we take a different path by directly combining collaborative filtering signals with content information through collaborative alignment. This innovative approach significantly enhances recommendation performance by seamlessly blending collaborative and content-based information.


\section{Conclusion}
In this paper, we study the collaborative alignment problem to bridge the gap between collaborative filtering and the pre-trained language model. Taking advantage of the pre-trained language model, we first obtain the item semanticized embedding with rich semantic information. Then, we propose CARec to encode user embedding into the item's semantic embedding space based on the collaborative signal. CARec treats users and items in different roles to better utilize both the collaborative filtering signal and the rich semantic information on items. Experiments on three real-world datasets under both warm and cold settings show our proposed CARec surpasses current state-of-the-art methods. Our case study on learned embedding space highlights that CARec can keep the semantic information on the semantic embedding space from pre-trained language model.

\bibliographystyle{ACM-Reference-Format}
\bibliography{sample-base}

\appendix

\section{appendix}
\subsection{Experimental Setup} \label{sec:experiment_setup}


\subsubsection{Compared Methods}
We compare the proposed approach with the following baseline methods:
\begin{itemize}[leftmargin=*]
    \item \textbf{NeuMF}~\cite{neumf} is a neural network-enhanced matrix factorization model that replaces the conventional dot product with a multi-layer perceptron (MLP) to capture more nuanced user-item interactions.
    
    \item \textbf{DirectAU}~\cite{directau} introduces an innovative loss function that evaluates representation quality in collaborative filtering (CF) based on alignment and uniformity within the hypersphere. In our implementation, we employ the alignment and uniformity loss, updating only the student role.
    
    \item \textbf{LightGCN}~\cite{lightgcn} represents a state-of-the-art recommendation algorithm grounded on Graph Convolutional Networks (GCN)~\cite{gcn}. It enhances performance by omitting feature transformations and nonlinear activations.
    
    \item \textbf{SimpleX}~\cite{simplex} proposes an easy-to-understand model with a unique loss function that incorporates a larger set of negative samples and employs a threshold to eliminate less informative ones. It also utilizes relative weights to balance the contributions of positive-sample and negative-sample losses.
    
    \item \textbf{NCL}~\cite{ncl} offers a neighborhood-enriched contrastive learning framework tailored for graph collaborative filtering. It explicitly captures both structural and semantic neighbors as objects for contrastive learning.
    
    \item \textbf{Wide\&Deep}~\cite{wide&deep} is a context-aware recommendation model that trains both wide linear models and deep neural networks concurrently, aiming to synergize the benefits of both memorization and generalization in RecSys.
    
    \item \textbf{DCNV2}~\cite{dcnv2} is another context-aware recommendation model that enhances the expressive power of Deep \& Cross Networks (DCN) by extending the original weight vector into a matrix. It also incorporates Mixture-of-Experts (MoE) and low-rank techniques to reduce computational overhead.

    \item \textbf{DroupoutNet}~\cite{droupoutnet} employs a dropout operation during training, randomly discarding portions of the collaborative embeddings. This technique effectively addresses the cold-start problem by enhancing the model's robustness.

    \item \textbf{Heater}~\cite{heater} utilizes the sum squared error (SSE) loss to model collaborative embeddings based on content information. Additionally, it employs a randomized training approach to further enhance the model's effectiveness.

\end{itemize}

\subsubsection{Evaluation Settings}
We evaluate our model's recommendation performance using commonly employed metrics in the field of Recommender Systems (RecSys): Recall@K and NDCG@K. By default, we set the values of K to 10 and 50. The reported results are based on the average scores across all users in the test set. These metrics consider the rankings of items that users have not interacted with yet. In line with established practices~\cite{lightgcn, neumf}, we utilize a full-ranking technique, which involves ranking all non-interacted items for each user. To assess the model's performance in a cold setting, we follow the procedures outlined in previous studies~\cite{multi-task-item-attribute-graph, droupoutnet, heater, clcrec}. In the Cold-start scenario, we identify cold-start items by removing all training interactions for randomly selected subsets of items.

\subsection{Implementation Details}
\label{sec:implementation_details}
We implement CARec and other baseline models using the open-source recommendation library, RecBole\footnote{https://recbole.io/docs/index.html}~\cite{recbole[1.1.1]}. For the sake of a fair comparison, we employ the Adam optimizer across all methods and conduct meticulous hyperparameter tuning. The batch size is configured at 1,024, and we implement early stopping with a patience setting of 30 epochs to mitigate overfitting, using NDCG@10 as our evaluation metric. Item text embeddings are generated utilizing a pretrained \textit{Instructor-xl} model\footnote{https://huggingface.co/hkunlp/instructor-xl}~\cite{instructor}. The dimensions for both user and item embeddings are fixed at 768. The instruction we use is "Represent the Amazon title:". 

For residual hyperparameters, we employ a grid search strategy to identify optimal settings. Specifically, the learning rate is explored within the set \{0.0001, 0.001, 0.01\}, and the weight decay coefficient is tuned among \{$1e^{-4}$, $1e^{-5}$, $1e^{-6}$\}. For graph-based models, the number of layers is evaluated over \{1, 2, 3\}. For content-based models, item semantic representations serve as item features. Within the MLP in CARec, we explore configurations with layer counts in \{1, 2, 3\}, hidden dimensions in \{384, 768, 1536\}, and dropout rates in \{0.2, 0.5\}.

\balance

\subsection{In-depth Discussions on CARec}

\subsubsection{Why Did We Choose Alignment and Uniformity Loss Over CE or BPR?} 

Our choice of using uniformity and alignment losses over cross-entropy (CE) or Bayesian Personalized Ranking (BPR) loss is driven by the specific objective of our model: to align the user representation space with the item semantic representation space while preserving semantic item representations.

Alignment loss directly aligns with our motivation by encouraging the user and item representations to be close in the same representation space. However, CE loss, while sharing a similar objective of alignment, may not guarantee that the aligned representations are adequately spaced apart. This distinction makes alignment loss a more suitable choice for our task. Moreover, we incorporate uniformity loss as it serves to prevent representations from becoming excessively close, effectively maintaining a balanced distribution. This is important to avoid over-concentration and promote diversity in the representation space, which aligns with our goal of preserving the richness of semantic semantics.

On the other hand, Bayesian Personalized Ranking (BPR) loss is designed to optimize item rankings for individual users, enhancing personalized recommendations. While it is valuable in its own right, it does not align with our specific motivation of preserving semantic item representations through the alignment of user and item spaces. Therefore, BPR loss is not aligned with the objectives of our model.

\subsection{Time Complexity}
While CARec involves training in two phases, the increase in training time is not substantial for two key reasons:

\begin{itemize}[leftmargin=*]
    \item In terms of time complexity, the most computationally intensive part of CARec is the uniformity loss. This loss component calculates similarity between each pair of items and users, resulting in a time complexity of $\mathcal{O}(|\mathcal{U}|^2+|\mathcal{I}|^2\times d)$. Fortunately, during each stage, we only need to compute the representation pairs that have been trained. For instance, during the semantic aligning phase, we only calculate user-user pairs, and during the collaborative refining phase, we only compute item-item pairs. This reduces the time complexity to $\mathcal{O}(max(|\mathcal{U}|^2,|\mathcal{I}|^2)\times d)$.
    \item The collaborative refining phase converges quickly. Once user representations have been aligned into the item semantic representation space, the item representation learning from user presentations becomes more efficient. For example, in our experiments, we trained for 25 epochs in the Electronic dataset, 57 epochs in the Office Products dataset, and 75 epochs in the Grocery and Gourmet Food dataset during the collaborative refining phase.
\end{itemize}

These factors collectively ensure that CARec's training time remains manageable even with the inclusion of additional phases.

\subsection{Limitation and Future Works}
This study is merely a preliminary exploration of collaborative alignment and carries several limitations. Firstly, we focused on recommendation scenarios involving only semantic item information, leaving room for future research to consider other context sources such as images and videos. Secondly, we have yet to fully explore scenarios involving user semantic information paired with item IDs, as well as those involving user semantic information matched with item semantic information, as each of these situations may entail distinct user and item roles. Lastly, the applicability of PLM-generated semantic embeddings remains an open question, especially when applied to different types of products such as books, music, movies, or restaurants. Further research is needed to uncover the utility of PLMs in these domains.

In our future work, one intriguing avenue of research revolves around preserving the semantic representations of multiple domains while effectively integrating collaborative filtering signals. To elaborate, different domains possess distinct semantic features, and pre-trained language models (PLMs) may encode these features into separate representation spaces. This diversity can pose challenges when attempting to incorporate them into a unified model. For instance, consider the domains of electronics and fashion. PLMs may encode the semantic features of electronic products differently from those of fashion items. Integrating recommendations across these domains necessitates bridging these distinct representation spaces, which presents a complex and interesting challenge.

Another compelling direction for future research centers on determining when a user, in their role as a learner, has acquired sufficient knowledge to transition to the role of tutor and vice versa for items. Currently, we rely on a straightforward validity score as an indicator, which may lead to premature or overly conservative transitions. Developing a novel indicator that more accurately gauges the readiness for role-switching could lead to substantial improvements in the model's performance. This indicator should strike a balance between stability and adaptability to optimize the learning process.

\end{document}